\documentclass[aps,prb,preprint,notitlepage]{revtex4-1}
\usepackage{amsmath}
\usepackage{graphicx}
\usepackage{amssymb}

\usepackage{gt17}
\usepackage{color} 
\usepackage{graphicx}
\usepackage{bm}
\usepackage{amsmath}
\begin{document}
\title{
Hydrodynamic theory of chiral angular momentum generation in metals 
}
\author{Hiroshi Funaki} 
\affiliation{5-27-5 Higashiohizumi, Nerima, Tokyo, 178-0063 Japan}
\author{Gen Tatara} 
\affiliation{RIKEN Center for Emergent Matter Science (CEMS), 
and RIKEN Cluster for Pioneering Research (CPR), 
2-1 Hirosawa, Wako, Saitama, 351-0198 Japan}

\date{\today}

\begin{abstract}
 We present a hydrodynamic theory to describe a chiral electron system with a Weyl spin-orbit interaction on a field-theoretic basis. 
 Evaluating the momentum flux density as a linear response to a driving electric field, we derive an equation of motion for the orbital angular momentum. 
 It is shown that the chiral nature leads to a dynamic bulk angular momentum generation by inducing a global torque as a response to the applied field.
 The steady state angular momentum is calculated taking account of rotational viscosity.  
\end{abstract}

\maketitle
Hydrodynamic theory is essential in describing fluids at macroscopic scales.
Conventional fluids are described by momentum flux tensor $\pi_{ij}$ symmetric with respect to the direction of momentum $i$ and flow $j$ \cite{LandauLifshitz-FluidMechanics}. 
Recently, fluids consisting of particles with internal degrees of freedom such as polar and chiral natures are drawing considerable attention, as such symmetries lead to unique fluids (chiral fluids and polar fluids) with different symmetries of momentum flux tensor.
For instance, chiral fluids have antisymmetric components of the tensor, which govern angular momentum dynamics of the fluids, playing essential roles  in many biological phenomena \cite{Kruse05,Furthauer12,Markovich19}. 

Hydrodymanic theory is based on equations representing continuity of densities of fluid,  momentum and energy at macroscopic scales. The coefficients in the equations are generally treated as phenomenological parameters determined by the symmetry of the system. 
Even microscopic quantum objects such as electrons in conducting solids are treated as a fluid when viewed at macroscopic scales.
Hydrodymanic approach has in fact been applied to study electron transport properties of solids. Most of the studies focused on viscosity induced by electron interactions  \cite{Conti99,Principi16,Lucas18}. Recently an anomalous transport in chiral electron system was discussed \cite{Toshio20}. A hydrodynamic viewpoint combined with linear response theory was applied to discuss vorticity dynamics \cite{Tserkovnyak19}.

For hydrodynamic equations describing macro(or meso)scopic fluid dynamics of electrons, classical treatments are generally sufficient  and thus most studies are based on a semiclassical arguments such as the Boltzmann equation and Berry's phase representation \cite{Chang96}. 
Applying hydrodynamic theory to quantum objects in solids has an advantage that the coefficients characterising the fluid can in principle be microscopically determined by use of a linear response theory with systematic approximations \cite{Conti99,Gao10,Principi16}.
In such microscopic methods, continuity equation for the density is automatically guaranteed by keeping the gauge invariance and only the equation for the momentum density needs to be considered. 

The aim of this paper is to construct a hydrodynamic theory for describing  chiral electron systems in solid on a field-theoretic basis.
The chiral nature is introduced by a three-dimensional spin-orbit interaction linear in the wave vector $\kv$, called the Weyl spin-orbit interaction.
Based on a quantum field representation, we derive an equation of motion for momentum operator and identify the momentum flux tensor as a field operator. 
The expectation value of the tensor is evaluated by use of non-equilibrium Green's functions as a linear response to a driving force, the applied electric field $\Ev$.
The coarse-grained equation obtained this way is the hydrodynamic equation
with each term evaluated fully quantum mechanically  without assuming classical transport equation. 
Based on the equation, we demonstrate that a field applied to a chiral fluid induces a bulk torque, resulting in a bulk orbital angular momentum generation.

The field Hamiltonian we consider is ($\hat{\ }$ denotes field operator) 
\begin{align}
 \hat{H}&=\sumkv \hat{c}_\kv^\dagger \lt(\ekv+\gammav_\kv\cdot\sigmav \rt) \hat{c}_\kv
\end{align}
where $\hat{c}$ and $\hat{c}^\dagger$ are conduction electron field operators with two spin components,  $\sigmav$ is a vector of Pauli matrices, $\ekv\equiv \frac{k^2}{2m}-\ef$ is the free electron energy ($\ef$ is the Fermi energy, $m$ is the electron mass).
A vector $\gammav_\kv\equiv \lambda \kv$ represents the Weyl spin-orbit interaction with a coupling constant $\lambda$. 
The electron velocity operator in the momentum space is
\begin{align}
 \hat{v}_i &= \frac{k_i}{m}+ a_i^\alpha\sigma_\alpha 
\end{align}
where 
\begin{align}
 a_i^\alpha=\frac{d \gamma_\kv^\alpha}{dk_i}
\end{align}
is the anomalous velocity due to the spin-orbit interaction. 

Our hydrodynamic equation is derived by calculating $\dot{\hat{\pv}}$, the time-derivative of the momentum operator, which reads $\dot{\hat{p}}_i=\nabla_{j}\hat{\pi}_{ij}$, where 
$\hat{\pi}_{ij}\equiv \hat{c}^\dagger \hat{p}_i \hat{v}_j \hat{c}$ is the momentum flux tensor operator. 
We thus obtain a coarse-grained hydrodynamic equation 
 \begin{align}
 \dot{{p}}_i=\nabla_{j}{\pi}_{ij} \label{hydro1}
 \end{align}
for  expectation values ${p}_i$ and ${\pi}_{ij}$ of $\hat{p}_i$ and $\hat{\pi}_{ij}$, respectively. 
Equation (\ref{hydro1}) obtained as a field average applies to length scales much larger than the electron mean free path. 
This simple equation is sufficient in most cases to describe linear regime of the fluid. 
In conventional fluids, the momentum flux tensor ${\pi}_{ij}$ has a diagonal part $-P\delta_{ij}$  representing a pressure $P$, and a symmetric part $\eta_{\rm s}(\nabla_i v_j+\nabla_j v_i)$ linear in both velocity $v_i$ and the spatial derivative, representing a viscosity $\eta_{\rm s}$
(See Eq. (\ref{Petas})) \cite{LandauLifshitz-FluidMechanics}.

For fluid dynamics with local velocity $\vv$, external driving force is necessary. 
We consider an electric field, represented by a vector potential $\Av$ as $\Ev=-\dot{\Av}$. 
Considering a linear response regime,  a local fluid velocity $\vv$ arises as a linear function of the local electric field.
As the system of our interest are in the weakly driven fluid regime, 
quadratic terms in $\vv$ commonly argued in the standard hydrodynamic equations are neglected.  

Considering a static limit, the momentum flux tensor at the linear order reads  $ \pi_{ij}=\pi_{ijk}E_{k}$, where the response function is 
$\pi_{ijk}(\qv)= \lim_{\omega_0\ra0}\frac{i}{\omega_0}\sumom \frac{e}{V}\sumkv  \tr[k_i \hat{v}_j G_{\omega,\kv} \hat{v}_k G_{\omega+\omega_0,\kv+{\qv}}]^<$, where $G_{\omega,\kv}$ is the path-ordered Green's function \cite{Rammer07} with angular frequency $\omega$ and wave vector $\kv$, $^<$ denotes the lesser component, 
$\tr$ is a trace over the spin indices, $e$ is the electron charge, and $V$ is the system volume. 
The wave vector and angular frequency of the external field are $\qv$ and $\omega_0$, respectively, and  the static limit  $\omega_0\ra0$ is taken at the end of the calculation. 
Focusing on the dynamically-driven flow, the dominant excited contributions  at low temperatures reads 
\begin{align}
 \pi_{ijk}(\qv)&=  \frac{e}{V} \sumkv 
 \tr[k_i \hat{v}_j G_{\kv}^\ret \hat{v}_k G_{\kv+\qv}^\adv]
 \label{pidef2}
\end{align}
where $G^\lambda_{\kv}\equiv G^\lambda_{\omega=0,\kv}$  ($\lambda=\adv,\ret$), $G^\ret_{\kv}=\frac{1}{-\epsilon_k-\gammav_k\cdot\sigmav + \frac{i}{2\tau} }$ and $ G_{\kv}^\adv = (G_{\kv}^\ret)^* $ 
are the retarded and advanced Green's function, respectively.
We take account of finite elastic electron lifetime $\tau$ as an imaginary part of the energy in the Green's functions.

Evaluating the trace over spin, we notice a striking fact that an antisymmetric uniform ($q=0$) component arises from the noncommutative two anomalous velocities proportional to 
$\av_j\times \av_k$ \cite{Chiral20}.
We have 
\begin{align}
 \pi_{ijk}(0) &= -\epsilon_{ijk}c
\end{align}
where
\begin{align}
 c &= \frac{4}{3}\lambda^2  \frac{e}{V}\sumkv 
 k  \Im(f_{\kv}^\ret {h}_{\kv}^{\adv} )
 \label{piq02}
\end{align}
is a bulk chiral coefficient ($k\equiv|\kv|$) \cite{Chiral20}. 
Here 
$f_\kv^\lambda = \frac{1}{2}\sum_{\sigma=\pm}g_{\kv\sigma}^\lambda$,  
$h_{\kv}^\lambda = \frac{1}{2} \sum_{\sigma=\pm}\sigma g_{\kv\sigma}^\lambda$, 
$g_{\kv\sigma}^\ret \equiv  \frac{1}{-\epsilon_k-\sigma \gamma_k +\frac{i}{2\tau} } $
is a spin-diagonalized Green's function, $\gamma_\kv\equiv|\gammav_{\kv}|$.

The contributions to $\pi_{ijk}$ linear in $\qv$ represent viscosity and pressure. 
They 
consist of symmetric and antisymmetric parts 
($ \pi_{ijk}^{(1)} =  \pi_{ijk}^{(1){\rm s}} + \pi_{ijk}^{(1){\rm a}}$),   
\begin{align}
 \pi_{ijk}^{(1){\rm s}} &= -i P q_k\delta_{ij} +i\eta_{\rm s}(q_i\delta_{jk}+q_j\delta_{ik})
 \label{Petas}
\end{align}
and 
\begin{align}
 \pi_{ijk}^{(1){\rm a}} &= -i\eta_{\rm a}(q_i\delta_{jk}-q_j\delta_{ik})
\end{align}
respectively.
The antisymmetric  viscosity constant is (explicit expressions for $P$ and $\eta_{\rm s}$ are suppressed) \cite{Chiral20}
\begin{align}
\eta_{\rm a} &=
-\frac{\lambda^2}{3m} \frac{e}{V} \Im \sumkv  k^2   
 \biggl[
\frac{1}{\gamma_\kv} f_\kv^{\ret} {h}_\kv^{\adv} 
+ f_\kv^{\ret} f_\kv^{\adv(2)} - {h}_\kv^{\ret} {h}_\kv^{\adv(2)}
+\frac{m\lambda}{k}( f_\kv^{\ret} {h}_\kv^{\adv(2)} - {h}_\kv^{\ret} f_\kv^{\adv(2)}) 
  \biggr]
 \label{pilineara}
\end{align}
where $f_{\kv}^{\lambda(2)}\equiv \frac{1}{2}\sum_{\sigma=\pm}(g_{\kv\sigma}^\lambda)^2$ and similarly for $h_\kv^{\lambda(2)}$.
Summing over $\kv$, the bulk chiral coefficient is 
\begin{align}
 c & 
= \frac{\pi}{3}\frac{e\lambda^2}{a^3} \sum_{\sigma} \frac{\nu_\sigma \gamma_\sigma k_\sigma}{(\gamma_\sigma)^2+\frac{1}{4\tau^2}} 
 \label{cresult}
\end{align}
where $\nu_\sigma$ and $k_\sigma$ are is spin-resolved electron density of states per site and wave vector, respectively, $\gamma_\sigma\equiv \gamma_{k_\sigma}$.
In the clean limit, $\gamma_\sigma\tau \gg 1$, 
$c\simeq \frac{\pi}{3}\frac{e\lambda^2}{a^3}\sum_{\sigma}\frac{\nu_\sigma k_\sigma}{\gamma_\sigma}$, while 
$c \simeq\frac{4\pi}{3}\frac{e\lambda^2}{a^3} \sum_{\sigma} {\nu_\sigma \gamma_\sigma k_\sigma} \tau^2$ 
in the disordered limit $\gamma_\sigma\tau \ll1$. \cite{Chiral20} 
The rotational viscosity is similarly evaluated to obtain  
$\eta_{\rm a}/c \simeq (\kf)^{-1}$, where $\kf$ is the Fermi energy, for both clean and dirty limits.

The hydrodynamic equation for a chiral system we obtained is therefore 
\begin{align}
 \dot{\pv} &= -c(\nabla\times\Ev) -(P-\eta_{\rm s})\nabla(\nabla\cdot\Ev)
  +\eta_{\rm s}\nabla^2 \Ev
  +\eta_{\rm a}[\nabla\times(\nabla \times \Ev)]  
  \label{hydroresult}
\end{align}
The driving field $\Ev$ is written in terms of a current $\jv$ using  a local identity $\jv=\sigma_{\rm e}\Ev$, where $\sigma_{\rm e}$ is local conductivity tensor, resulting in a standard hydrodynamic equation relating the force and velocity. 
This equation for chiral fluid, obtained microscopically for the Weyl electron system, applies to general chiral matter. 
In fact, the same equation has been argued on the phenomenological basis in the context of chiral active matter for describing biological phenomena \cite{Furthauer12,Markovich19}.
The global chiral term $c$ was argued also in noncentrosymmetric metals and was shown based on a semiclassical argument \cite{Toshio20}.

Chiral effects in Eq. (\ref{hydroresult}) arise from rotational flow or vorticity, 
$\nabla\times\jv$. 
The term $\eta_{\rm a}$ represents the rotational viscosity, while the bulk chiral coefficient, $c$, represents a global angular momentum generation.  
The roles of chiral terms become clear focusing on the total angular momentum of the system,
$\Lv=\intr (\rv\times\pv)$. 
Here $\rv$ in the present fluid description is a coordinate (and not a dynamic variable) and the spatial integral is over the system with a finite volume, $V$. 
The torque (time-derivative of $\Lv$) on the system is 
$ \dot{L}_i = \epsilon_{ijk} \intr r_j \nabla_l \pi_{kl}$. 
As the current density and thus the field $\Ev$ and $\pi_{kl}$ vanish at the surface of the system, we use the integral by parts to obtain 
$ \dot{L}_i =  \epsilon_{ijk} \intr  \pi^{\rm a}_{jk}= \intr [ 
 -c E_i +\eta_{\rm a}(\nabla\times\Ev)_i ]  $, 
namely, only the antisymmetric part of the momentum flux density  contributes to the total torque.
Considering the case of a uniform applied field, the rotational contribution arises as a result of induced current $\jv$ in the plane perpendicular to the applied field. 
We thus use $\Ev=\jv/\sigma_{\rm e}$, where $\sigma_{\rm e}$ is approximated as diagonal for simplicity, to rewrite the rotation term.  
The equation representing chiral angular momentum generation due to a driving field is therefore  
\begin{align}
 \dot{\Lv} &= \intr \biggl[ 
 -c \Ev +\frac{\eta_{\rm a}}{\sigma_{\rm e}} (\nabla\times\jv) \biggr]  
 \label{LbyE}
\end{align}
This equation applies to nonelectron cases if the field $\Ev$ is to be replaced by  a corresponding driving force.
The steady state is the one when the bulk chiral torque ($c$ term) is balanced by the rotational viscosity.

Let us consider a cylinder along the $z$ axis with length $l_z$ and radius $R$. When an electric field is applied along the cylinder, an in-plane current perpendicular to the field is induced due to a chiral effect represented by antisymmetric components of the conductivity tensor \cite{Chiral20}. 
Due to the rotational viscosity, the system reaches a steady state when the rotation is uniform with a constant angular frequency $\Omega$ determined by Eq. (\ref{LbyE}). In this steady state, the local current density at $\rv=(x,y)$ is   $\jv=en\Omega(\hat{\zv}\times\rv)$ ($n$ is the electron density), and local vortex density is uniform, 
$(\nabla\times\jv)_z=2en\Omega$. 
The steady condition, 
\begin{align}
 c E_z = \frac{\eta_{\rm a}}{\sigma_{\rm e}} (\nabla\times\jv) 
 \label{eqcond}
\end{align}
is therefore  satisfied locally at the angular frequency of 
\begin{align}
 \Omega&=  \frac{eE\tau}{m}\frac{c}{\eta_{\rm a}} = \frac{eE}{2}\frac{\ell}{\hbar}
 \label{eqfreq}
\end{align}
where $\ell=\kf\tau/m$ is the elastic electron mean free path and we used $\sigma_{\rm e}=e^2 n\tau/m$ and $c/\eta_{\rm a}\sim \kf$ in the last equality.
The steady state angular momentum of the electrons is $L_e=I_e\Omega$, where $I_{e} = \frac{m}{2} NR^2$ is the moment of inertia of the electrons, $N\equiv V/a^3$ being the number of lattice sites ($a$ is the lattice constant). 
Electron angular momentum per site is thus 
\begin{align}
 \frac{L_e}{N} &=  eE\frac{m}{4\hbar} \ell R^2 
 \label{LzN}
\end{align}
A significant feature here is the fact it is proportional to the area of the cylinder. This is because the present chiral mechanism, Eq. (\ref{eqfreq}), induces local vorticity, which is proportional to the angular frequency of the entire rotation. 
The total angular momentum  is thus proportional to the area as a result of $I_e\propto R^2$. 
Equation (\ref{LzN}) would, however, breakdown for large systems of $R\gg \ell_{L}$, where $\ell_L$ is a relaxation length of angular momentum.

The enhancement in large systems does not apply to the spin density induced by the electric field (Edestein effect). 
The response function for the induced spin, discussed on the equal footing as the orbital angular momentum,  is 
$ s_{jk}(\qv)
 =\frac{e}{V}\sumkv 
 \tr[\sigma_j G_{\kv}^\ret \hat{v}_k G_{\kv+\qv}^\adv]$ \cite{Chiral20}.
A uniform component exists, which is diagonal; $s_{jk}=\delta_{jk}\kappa_s$, where 
\begin{align}
 \kappa_s
 &= 2 \frac{e}{V}\sumkv \Re \biggl[
 \frac{2k}{3m} f_{\kv}^\ret  {h}_{\kv}^\adv 
 +\lambda \lt(f_{\kv}^\ret  f_{\kv}^\adv -\frac{1}{3} {h}_{\kv}^\ret h_{\kv}^\adv \rt)\biggr]
\label{s2}
\end{align}
The induced spin density is (with $\hbar$ recovered) 
\begin{align}
 s&=\kappa_s E \simeq \frac{eE}{\hbar} \frac{\tau}{\kf}
\end{align}
The induced spin density is determined simply by microscopic material parameters.
The ratio of the chirally-induced orbital and spin angular momenta is significantly large for large systems; $\frac{L_e}{N}/s\simeq (R/a)^2$. 
Although both are angular momenta, the efficiency for electric generation is by order of magnitude larger for the orbital one compared to the spin because of Eq. (\ref{LzN}).  
This large pumped angular momentum by electron injection, combined with spin-orbit interaction, may account for the large spin polarization of the electrons transmitted through chiral materials (spin selectivity) \cite{Naaman15}. 

Let us look into numbers. 
If we apply a voltage of 1V along a cylinder of $l_z=1\mu$m, $r=100$nm, $eEa=3.2\times10^{-23}$J for $a=2$\AA.
The angular momentum in unit of $\hbar$ per site is 
$\frac{L_e}{\hbar N}\simeq \frac{eEa}{\ef} (\kf R)^2\frac{\ell}{a}=0.5\times 10^4$
for $\ell/a=100$ and $\kf\sim a^{-1}$ and $\ef=1$eV, while the induced spin $s$ is $4\times10^{-6}$ times smaller.

When the cylinder is free to rotate, the angular momentum of the electrons is distributed to the rotation of the whole system. 
The rotation angular frequency in this case is 
$L_{\rm tot}= \frac{m}{N_{\rm a}m_{\rm p}}L_{\rm e}$, where $m_{\rm p}$ and $N_{\rm a}$ are the proton mass and atomic number.
Considering the case of C atom ($N_{\rm a}=12$), the suppression factor is 
$\frac{m}{N_{\rm a}m_{\rm p}}=4.5\times 10^{-5}$. 
If the orbital angular momentum estimated above is fully transferred to a mechanical rotation, 
it corresponds to the angular frequency of the total system of  
$\Omega_{\rm tot}=\frac{m}{N_{\rm a}m_{\rm p}}\Omega \sim \frac{m}{N_{\rm a}m_{\rm p}}eEa\frac{\ell}{\hbar a}=7\times 10^{8}$ Hz. 

Our result, Eq. (\ref{eqfreq}), indicates that chiral systems work as rotation sensors detected electrically. 
The voltage at a steady angular frequency $\Omega$ is $V_E=E l_z=\frac{2\hbar}{e}\frac{l_z}{\ell}\Omega$, which in the above case is $V_E=0.7\times 10^{-13} \times \Omega$[Hz] V.
For a device of $l_z=1$mm, the magnitude is of the order of 0.1$\mu$V for $\Omega=1$kHz 
 and would be detectable.
 
As our results (Eq. (\ref{cresult}) and Ref. \cite{Chiral20}) indicates that the bulk  chiral coefficient and rotational viscosity have different behaviors in the clean ($\gamma_\sigma \tau \gg1$) and dirty ($\gamma_\sigma \tau \ll1$) limits. 
Experimental control of $\gamma_\sigma$ by tuning the Fermi level would be of interest.

The angular momentum generation effect suggests existene of chiral edge current. In fact, by calculating the electric conductivity tensor, we find a chiral contribution, antisymmetric and linear in $q$, 
$\sigma^{\rm ch}_{jk} =   \sigma_{\rm a}i\epsilon_{jkl}q_l $, 
where $\sigma_{\rm a}$ is a constant \cite{Chiral20}.
This contribution represents a chiral current circulating in the plane perpendicular to the applied field at the edge of the system, where the $\nabla\times \Ev$ is finite. 
This chiral current, argued in Ref. \cite{Toshio20}, is interpreted as another aspect of the bulk chiral angular momentum generation.

Electric generation of orbital angular momentum in chiral crystal was theoretically pointed out in Ref. \cite{Yoda18} (called the orbital Edelstein effect). 
A realization in honeycomb-lattice layers was argued. 
Their approach is based on a semiclassical Berry phase formula, and thus is an equilibrium contribution.  
Their result of the induced magnetization was proportional to elastic lifetime $\tau$, which agrees with  the present dynamic approach, Eq. (\ref{LzN}). 
A Berry phase interpretation to the chiral coefficient $c$ was argued in Ref.  \cite{Toshio20}.

Equation (\ref{LbyE}) indicates that the rotational viscosity contributes to the diamagnetic constant. 
In fact, using $\nabla \times \Ev=-\dot{\Bv}$, we obtain standard diamagntism relation,  
$ \Lv/V = -\eta_{\rm a}\Bv$ when a magnetic field $\Bv$ is applied without $\Ev$. 
Note that the diamagnetic effect represented by the viscosity is only the dynamic contribution, as it was derived by evaluating $\dot{\Lv}$. 

In the eletromagnetism  viewpoint, the bulk chiral contribution, $\dot{\Lv} \propto c\Ev$,   is a cross-correlational diagmagnetic response to the electric field.
Mathematically the effect arises from a non-commutative off-diagonal contribution in the response function \cite{Chiral20}. 
Obviously, inversion symmetry breaking (chiral nature) is essential; For example,   
$c$  vanishes for the case of the Rashba interaction, $\gammav_\kv=\alphav\times\kv$ ($\alphav$ is the Rashba vector) \cite{Chiral20}. 
The existence of spin or pseudospin (sublattice) in addition to the parity symmetry breaking is therefore essential for the bulk chiral effect. 
The requirement is common to chiral and topological effects, suggesting close relation between the two. 
This point is further discussed in the viewpoint of electromagnetic response.

The induced angular momentum $\Lv$ of charge leads to a magnetization, $\Mv=\frac{e}{2m}\Lv$. 
Our result, Eq. (\ref{LzN}), therefore indicates that a coupling between $\Ev$ and $\dot{\Bv}$ is induced by a chirality. 
This is consistent with a recent finding \cite{Kawaguchi18} that electromagnetism in chiral material contains an effective coupling proportional to the chirality order parameter, $C\equiv \Ev\cdot(\nabla\times \Ev)=-\Ev\cdot\dot{\Bv}$.
This is natural as the coupling $\Ev\cdot\dot{\Bv}$ is chiral in the sense that it breaks the parity invariance keeping the time-reversal invariance \cite{Barron12}.
In this electromagnetism viewpoint, our chiral angular momentum effects is closely related to topological effects. 
In fact, the effective Hamiltonian for the electromagnetism in topological insulators is proportional to $\Ev\cdot\Bv$, resulting in a direct coupling $\Mv\propto \Ev$ \cite{Qi11} or $\Lv \propto \Ev$.
Therefore our dynamic bulk chiral effect is a non-equilibrium counterparts of topological effects,  inducing $\dot{\Lv}$ instead of $\Lv$ itself.
Moreover, the dynamic chiral effect leads to a steady state with a finite angular momentum (Eq. (\ref{LzN})), i.e., it has a qualitatively the same effect as the topological one. 
The field-induced orbital magnetization observed in experiments is thus written as 
$\Mv= c_{\rm top}\Ev+c_{\rm ch}\Ev$, where $c_{\rm top}$ is a coefficient for the topological effect, and $c_{\rm ch}\propto c$ represents the present dynamic chiral effect. Being topological, $c_{\rm top}$ is a universal constant, while $c_{\rm ch}$ arising as a result of a relaxation due to a viscosity  is of the order of 
$c_{\rm ch}/c_{\rm top}\simeq \tau_{L} \varepsilon$, where $\tau_{L}$ and $\varepsilon$ represent the relaxation time for the angular momentum and typical energy scale, respectively. 
In the present simple model, $\varepsilon\sim \ef$ and $\tau_{L}\propto \tau$. For weak angular momentum relaxation, the ratio is much larger than unity, suggesting that topological contribution might be smeared out.

To conclude, we have constructed a hydrodynamic theory to describe chiral electron systems by a straightforward calculation on a quantum field-theoretical basis. 
Deriving the equation of motion for the orbital angular momentum, we discussed a bulk chiral angular momentum generation effect when an electric field is applied. 
The magnitude of the orbital angular momentum is of the orders of magnitude larger than the spin angular momentum. 
The hydrodynamic equation itself applies generally for any chiral fluids if the electric field is replaced by the corresponding driving field.

\acknowledgements
The authors thank S. Seki, Y. Fuseya, J. Kishine for valuable discussion. 
This study was supported by
a Grant-in-Aid for Scientific Research (B) (No. 17H02929) from the Japan Society for the Promotion of Science.

%

\end{document}